\documentclass[twocolumn,showpacs,aps,prl,superscriptaddress]{revtex4}

\usepackage{graphicx}
\usepackage{dcolumn}
\usepackage{amsmath}
\usepackage{epsfig}

\input pubboard/babarsym
\input Definitions

\newcommand{\BABARPubYear}    {06}
\newcommand{\BABARPubNumber}  {039}

\newcommand{\SLACPubNumber} {12072}

\def\figurebox#1#2#3{%
    \def\arg{#3}%
    \ifx\arg\empty
    {\hfill\vbox{\hsize#2\hrule\hbox to #2{\vrule\hfill\vbox to #1{\hsize#2\vfill}\vrule}\hrule}\hfill}%
    \else
    {\hfill\epsfbox{#3}\hfill}%
    \fi}

\begin{document}

\preprint{\babar-PUB-\BABARPubYear/\BABARPubNumber} 
\preprint{SLAC-PUB-\SLACPubNumber} 

\begin{flushleft}
\babar-PUB-\BABARPubYear/\BABARPubNumber\\
SLAC-PUB-\SLACPubNumber\\
\end{flushleft}

\title{
{\large \bf
Measurement of the $CP$ Asymmetry and Branching Fraction of \boldmath$B^0\to\rho^{0}K^0$} 
}

%
\author{B.~Aubert}
\author{R.~Barate}
\author{M.~Bona}
\author{D.~Boutigny}
\author{F.~Couderc}
\author{Y.~Karyotakis}
\author{J.~P.~Lees}
\author{V.~Poireau}
\author{V.~Tisserand}
\author{A.~Zghiche}
\affiliation{Laboratoire de Physique des Particules, F-74941 Annecy-le-Vieux, France }
% ...
\collaboration{The \babar\ Collaboration}
\noaffiliation

\begin{abstract}
We present a measurement of the branching fraction and time-dependent $CP$ asymmetry of $B^0\to\rho^{0}K^0$. 
The results are obtained from a data sample of $227\times10^6$ 
$\FourS \to B\Bbar$ decays  collected with the \babar\  detector at the \pep2 asymmetric-energy $B$~Factory at SLAC.
From a time-dependent maximum likelihood fit yielding $111\pm19$ signal events we find ${\cal B}(B^0\to\rho^{0}K^0)=(4.9\pm0.8\pm0.9)\times10^{-6}$, where the first error is statistical and the second systematic. We report the measurement of the $CP$ parameters $S_{\rho^{0}\KS}=0.20\pm{0.52}\pm{0.24}$ and $C_{\rho^{0}\KS}=0.64\pm{0.41}\pm{0.20}$. 
\end{abstract}


\maketitle

\label{sec:Introduction}

Decays of $B^0$ mesons to the $\rho^0 K^0$ final state are expected to be dominated by $b\rightarrow s$ penguin amplitudes.
Neglecting Cabibbo-Kobayashi-Maskawa (CKM) suppressed amplitudes, the mixing-induced \CP violation
parameter $S_{\rho^{0}\KS}$ should equal $\sin 2\beta$, which is well measured in
$B^0 \rightarrow J/ \psi K^0$ decays \cite{Aubert:2004zt}.
Within the Standard Model (SM), only small deviations from this prediction are expected \cite{Chiang:2003pm}. In the Standard Model , a single phase in the CKM matrix governs \CP violation \cite{CKM},
but if heavy non-SM particles appear in additional penguin diagrams, 
new \CP-violating phases could enter and $S_{\rho^{0}\KS}$ would not equal $\sin 2\beta$ \cite{NewPhys}. Observation of a significant discrepancy would be a clear signal of new physics.
\par
In this Letter we present the first observation of the decay $B^0\to\rho^0K^0$ and a measurement of the \CP-violating asymmetries $S_{\rho^{0}\KS}$ and $C_{\rho^{0}\KS}$ from a time-dependent maximum likelihood analysis. A non-zero value of $S_{\rho^{0}\KS}$ indicates $CP$ violation due to the interference between decays with and without mixing. Direct $CP$ violation leads to a non-zero value of $C_{\rho^{0}\KS}$.
We take a quasi-two-body (Q2B) approach, restricting ourselves to the region of the $B^0\to\pi^+\pi^-\KS$ 
Dalitz plot dominated by the $\rho^0$ and treating other $B^0\to\pi^+\pi^-\KS$ contributions as non-interfering background. The effects of interference with other resonances are estimated and taken as systematic uncertainties.

The data were collected with the \babar\ detector at the \pep2 asymmetric-energy $e^+e^-$
storage ring at SLAC. An integrated luminosity of $205\invfb$, corresponding to 
$227\times10^{6}$ $\B\Bbar$ pairs, was collected at the \FourS resonance
(center-of-mass (CM) energy $\sqrt{s}=10.56\gev$), and $16\invfb$ was collected about $40$ MeV
below the resonance (off-resonance data).  
The \babar\ detector is described
in detail elsewhere \cite{bib:babarNim}. Charged particles are detected and their 
momenta measured by the combination of a silicon vertex tracker (SVT), consisting of
five layers of double sided detectors, and a 40-layer central drift chamber (DCH), both 
operating in the 1.5 T magnetic field of a solenoid. Charged-particle identification
is provided by the average energy loss in the tracking devices and by an internally
reflecting ring-imaging Cherenkov detector (DIRC) covering the central region.

\label{sec:selection}
\par
We reconstruct $B^0\to \rho^{0} \KS$ candidates ($B^0_{\rm rec}$ in the 
following)  from combinations 
of $\rho^0$ and $\KS$ candidates, both reconstructed in their $\pi^+\pi^-$ decay mode.
For the $\pi^+\pi^-$ pair  from the $\rho^{0}$ candidate, 
we remove tracks identified as very likely to be 
electrons, kaons, or protons.  
The mass of the 
$\rho^{0}$ candidate is restricted to the interval $0.4<m(\pi^+\pi^-)<0.9\gevcc$.    
The $\KS$ candidate is required to have a mass within $13\mevcc$ of 
the nominal $\KS$ mass \cite{PDG2004} and a decay vertex separated from the $\rho^0$ decay vertex
 by at least three times the estimated seperation measurement uncertainty. In addition, the cosine 
of the angle in the lab frame 
between the $\KS$ flight direction and the vector between 
the $\rho^0$ decay vertex and the $\KS$ decay vertex must be greater than 0.995. 
Vetoes against $B^0\rightarrow D^+\pi^-$ and $B^0\rightarrow K^*\pi^-(K^*\
 \rightarrow \KS\pi^+)$ are imposed by requiring that the invariant masses of both $\KS\pi$ combinations are more than $0.055\gevcc$ and $0.040\gevcc$ from the $K^{*+}$ and $D^+$ masses  \cite{PDG2004} respectively.
To exclude events with poorly reconstructed vertices we require the estimated error on $\Delta{t}$ to be less than $2.5\ps$ and that $|\Delta{t}|$ must be less than $20\ps$, where $\Delta t$ is the proper time difference
between the decay of the reconstructed $B$ meson ($B^0_{\rm rec}$) and its 
unreconstructed partner ($B^0_{\rm tag}$), $t_{\rm rec}-t_{\rm tag}$. It is 
determined
 from the measured relative displacement of the two $B$-decay vertices and the known boost of the $e^+e^-$ system.
\par
Two kinematic variables are used to discriminate between signal and combinatorial background. 
The first is $\Delta{E}$, the difference between the measured CM energy of the $B$~candidate and $\sqrt{s}/2$, where $\sqrt{s}$ is the CM 
beam energy. 
The second is the beam-energy substituted mass 
$\mes\equiv\sqrt{(s/2+{\mathbf {p}}_i\cdot{\mathbf{p}}_B)^2/E_i^2-{\mathbf {p}}_B^2}$,
where the $B^0_{\rm rec}$ momentum ${\mathbf {p}}_B$ and the four-momentum of the 
initial $\Upsilon(4S)$ state ($E_i$, ${\mathbf {p}}_i$)
 are defined in the laboratory frame. We require $|\Delta{E}|<0.15\gev$ and $5.23<\mes<5.29\gevcc$.
\par
Continuum $e^+e^-\to q\bar{q}$ ($q = u,d,s,c$) events are the dominant 
background.  To enhance discrimination between signal and continuum, we 
use a neural network (NN) to combine five  variables: the 
cosine of the angle between the $B^0_{\rm rec}$ direction 
and the beam axis in the CM,
the cosine of the angle between the thrust axis of the
$B^0_{\rm rec}$  candidate
and the beam axis, the sum of momenta transverse to the direction of flight of 
the $B^0_{\rm rec}$, and the zeroth and second angular moments $L_{\rm 0,2}$
of the energy flow about the $B^0_{\rm rec}$ thrust axis.  The moments
are defined by $L_j=\sum_i \mathbf{p}_i \times |\cos{\theta_i}|^j$, 
where $\mathbf{p}_i$ is its momentum and $\theta_i$
is the angle with respect to the $B^0_{\rm rec}$ thrust axis of the track
or neutral cluster $i$ excluding the tracks that make up the 
$B^0_{\rm rec}$ candidate.
The NN is trained 
with off-resonance data and
Monte Carlo (MC)\cite{Agostinelli:2002hh} simulated signal events. 

\par
selected signal events are reconstructed incorrectly with low momentum tracks from the other $B$ meson being used to form the $\rho^0$ candidate.
In total, 20,073 events pass all selection criteria in the on-resonance sample.
\par

\label{sec:BBackground}
An unbinned extended maximum likelihood fit is used to
extract the $\rho^{0}\KS$ $CP$ asymmetry and branching fraction.
There are ten components in the fit: signal, continuum background and 
eight separate backgrounds from $B$ decays. 
Large samples of MC-simulated events are used to identify these specific $B$ backgrounds. 
Where an individual decay mode makes a significant 
contribution to the dataset (one or more events expected in the data) we include it as a separate contribution to the fit. 
Probability density functions (PDFs) are taken
from simulation with the expected number of $B$ background 
events fixed to values estimated from known branching fractions \cite{PDG2004}
and MC efficiencies (Table \ref{tab:bbkg}).
Where only upper limits are available, decay modes are not 
included in the default fit but are used in alternate fits to evaluate systematics.
\par Events from $B$ decays that do not come from individually significant channels are collected together into two ``bulk'' $B$ contributions to the fit ($B^0$ and $B^+$). 
The assumption is made that $\Bz\to{f_0(600)}\KS$ can be neglected, with support from \cite{bellekpipi,babarkpipi} which do not require this mode to describe $\Bp\to{K^+}\pi^+\pi^-$.

\par
\label{sec:themlfit}
\par
  The events in the data sample have their unreconstructed $B$s flavor-tagged as $B^0$ or $\Bzb$ with the method described in \cite{BReco}. Events are separated into four flavor-tagging categories and an ``untagged'' category, depending upon the method used to determine the flavor. Each category has a different expected purity and accuracy of tagging. The likelihood function for the $N_\cat$ candidates in flavor tagging category $k$ is

\begin{eqnarray}
\label{eq:pdfsum}
{\cal L}_k &=e^{-N^{\prime}_\cat}\!\prod_{\rm i=1}^{N_\cat}
		\bigg\{   N_{S}\epsilon_\cat\left[
                                (1-f^\cat_\textrm{MR}){\cal P}_{\rm i,\cat}^{S^\textrm{CR}} +
                                f^\cat_\textrm{MR}{\cal P}_{\rm i,\cat}^{S^\textrm{MR}}
                              \right]\nonumber\\
	&+ N_{\rm C,\cat} {\cal P}_{\rm i,\cat}^{C} + \sum_{\rm j=1}^{n_B} N_{\rm B,j} \epsilon_{\rm j,\cat}{\cal P}^{\B}_{\rm ij, \cat}
	\bigg\},
\end{eqnarray}

\noindent where $N^{\prime}_\cat$ is the sum of the signal and background yields for events tagged in category $\cat$,
 $N_S$ is the number of 
$\rho^{0}\KS$ signal events in the sample, $\epsilon_\cat$ is the 
fraction of signal events tagged in category $\cat$, $f^\cat_\textrm{MR}$
is the fraction of mis-reconstructed (MR) signal events in tagging category
$\cat$ and the superscript CR implies correctly reconstructed signal. 
$N_{\rm C,\cat}$ 
is the number of continuum background events  that are tagged in 
category~$\cat$, and $N_{\rm B,j}\epsilon_{\rm j,\cat}$ is the number of
$B$-background events of class $j$ that are tagged in category~$\cat$. 
The \B-background event yields are fixed in the default fit to values shown in
Table
\ref{tab:bbkg}. The values $\epsilon_k$ and $f^k$ are determined from MC for $B$-backgrounds and from a sample of $B$ decays of known flavor for signal.
The total likelihood 
${\cal L}$ is the product of the likelihoods for each tagging category.

\begin{table}
\begin{tabular}{lr@{$\pm$}l}
 \hline \hline
Background Mode &\multicolumn{2}{c}{$N_{\rm expected}$} \\
 \hline
$\rm Bulk~ \it B^+$ & $197$&$98$ \\
$\rm Bulk~ \it B^0$ & $197$&$98$ \\
$B^0 \rightarrow D^+ \pi^-$ & $40$&$6$ \\
$B^0 \rightarrow \eta'\KS$  & $34$&$5$ \\
$B^0 \rightarrow f_0(980)\KS $ & $22$&$4$ \\
$B^0 \rightarrow K^*_0(1430)^+ \pi^-$ & $7$&$1$ \\
$B^0 \rightarrow \rho^0 K^{*0}$ & $3$&$3$ \\
$B^0 \rightarrow (\KS \pi^+ \pi^-)_{\rm NR} $ &  $2$&$1$ \\
\hline \hline
\end{tabular}
\caption{Expected number of events from each $B$ background source.}
\label{tab:bbkg}
\end{table}

Each signal and background PDF is defined as:
$  {\cal P}_\cat = 
	{\cal P}(\mes)\cdot 
	{\cal P}(\de) \cdot
 	{\cal P}_\cat(\NN) \cdot 
	{\cal P}(\cos\theta_{\pi^+}) \cdot
	{\cal P}(\Delta t) \cdot
	{\cal P}(m_{\pi^+\pi^-})
$ 
where \mes, \de, \NN, $m(\pi^+\pi^-)$ are the variables described previously, and $\cos\theta_{\pi^+}$ is the angle between the $\KS$ and the $\pi^+$ from the $\rho^0$ in the $\rho^0$ meson's center-of-mass frame. 

The $\Delta t$ PDF for signal events is defined as
\begin{eqnarray}
{\cal P}\left( \Delta t \right)=&\frac{e^{-|\Delta t|/\tau_B}}{4 \tau_{B}} \times \nonumber\\
&\left[ 1+\frac{\Delta D}{2}+q\langle D \rangle \left( S_{\rho^{0}\KS} \sin(\Delta m_d \Delta t) -\right.\right. \nonumber \\
&  \left. \left. C_{\rho^{0}\KS} \cos(\Delta m_d \Delta t)\right)\right] \otimes R_{\rm sig}(\Delta t, \sigma_{\Delta t}),
\label{eqn:deltat} 
\end{eqnarray}
\noindent
where $\tau_B$ and $\Delta m_d$ are the average lifetime and
eigenstate mass difference of the neutral $B$ meson, $q = +1~(-1)$
when $B^0_{\rm rec} = \Bz~(\Bzb)$, $\langle D \rangle$ describes the
dilution effect from imperfect flavor tagging, and $\Delta D$ is the
difference in this dilution between $\Bz$ and $\Bzb$ tags.
This formalism is found to effectively describe both correctly and incorrectly reconstructed signal.
$\langle D \rangle$, $\Delta D$ and the $\Delta t$ resolution function, $R_{\rm sig}(\Delta t, \sigma_{\Delta t})$, have parameters fixed to values taken from
a sample where $B$s of known flavor can be reconstructed \cite{BReco}. ``Untagged'' events have a $\langle D \rangle$ of 0, reflecting the lack of tag information.

\begin{figure}
  \centering
  \begin{tabular}{cc}
    \begin{minipage}{1.72in}
      \centering 
      \includegraphics[height=1.7in,width=1.7in]
		      {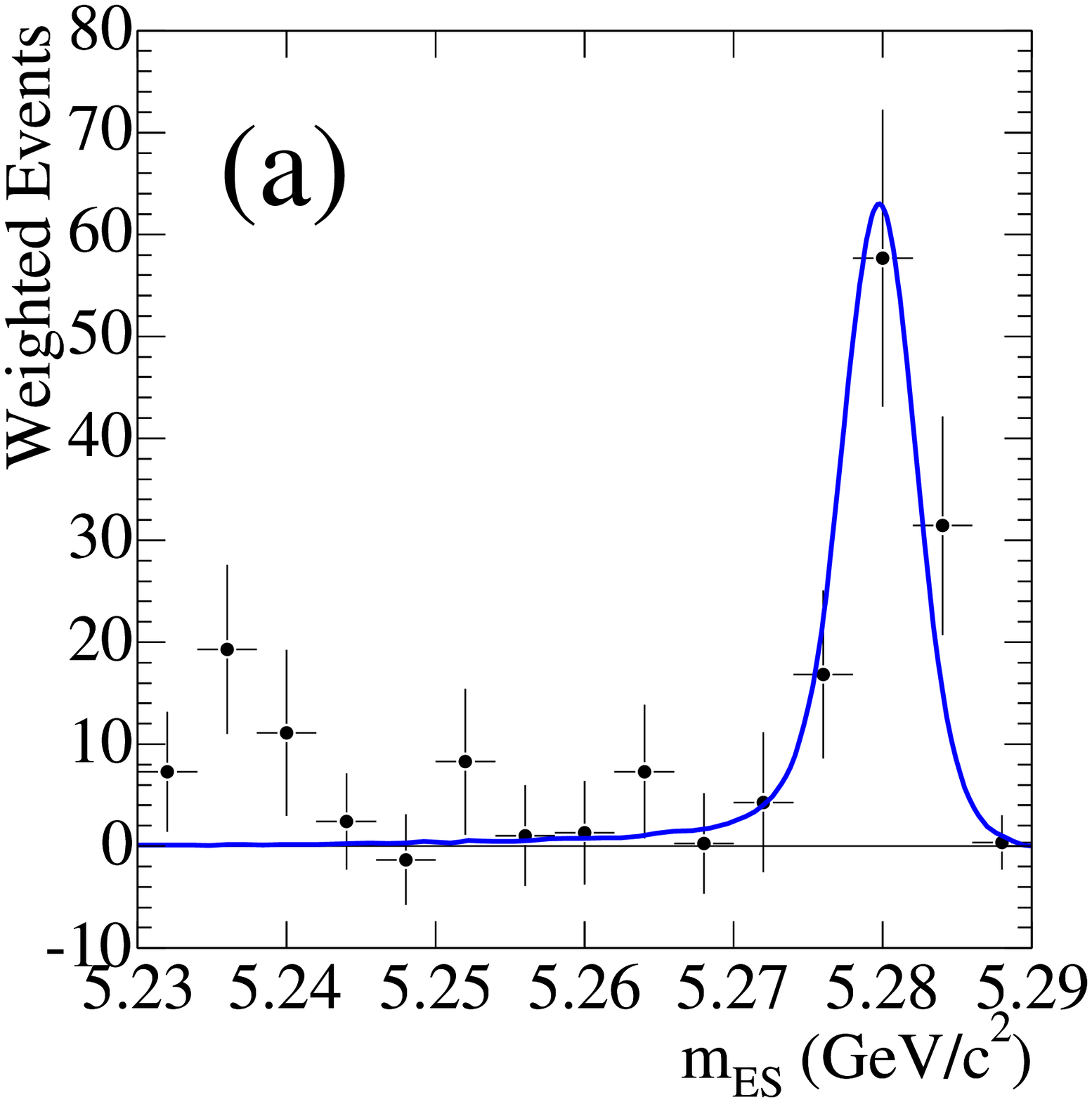}
    \end{minipage}
    &
    \begin{minipage}{1.75in}
      \centering
      \includegraphics[height=1.7in,width=1.7in]
		      {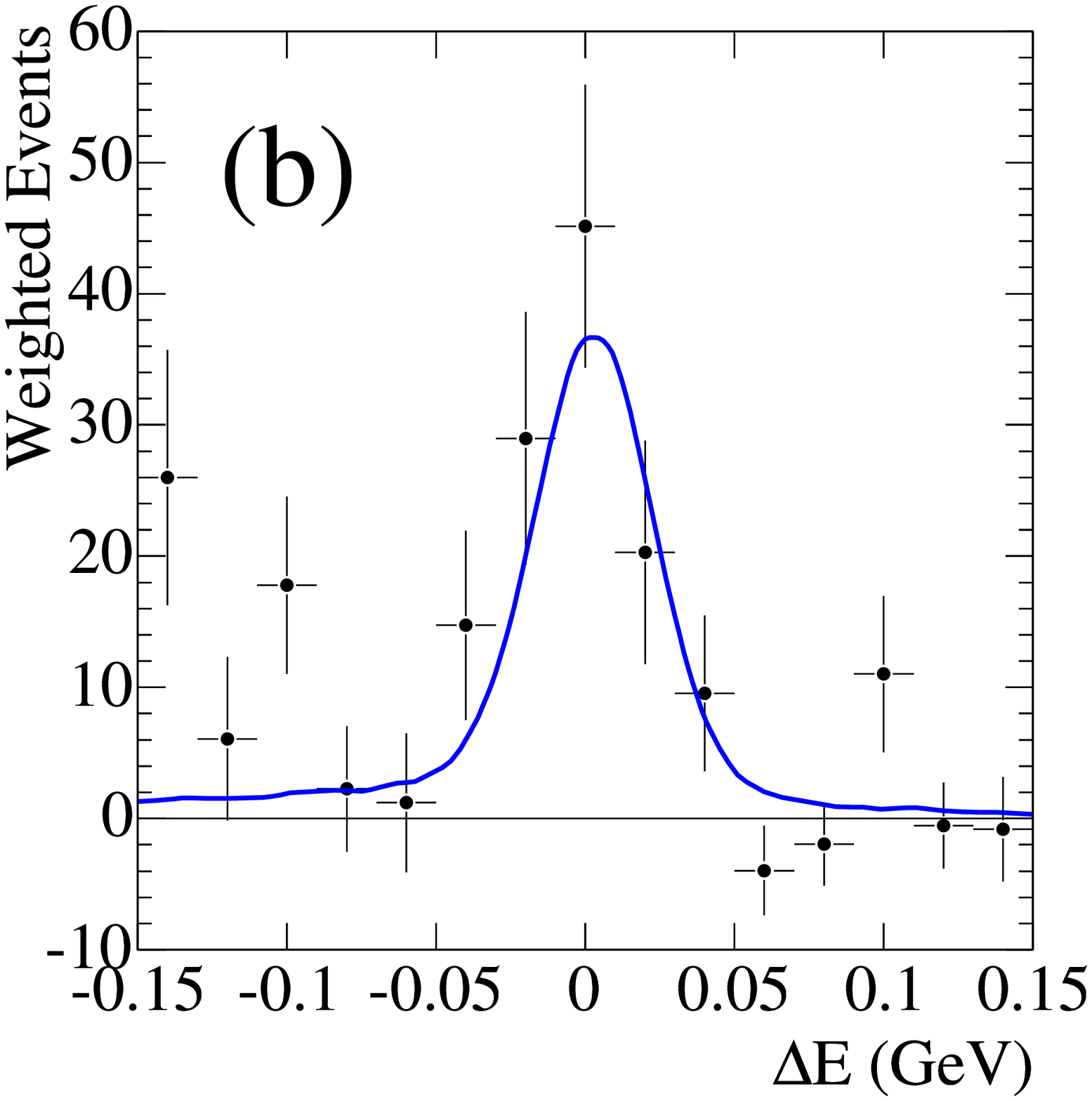}
    \end{minipage}
  \end{tabular}
  \centering
  \begin{tabular}{cc}
    \begin{minipage}{1.72in}
      \centering
      \includegraphics[height=1.7in,width=1.7in]
		      {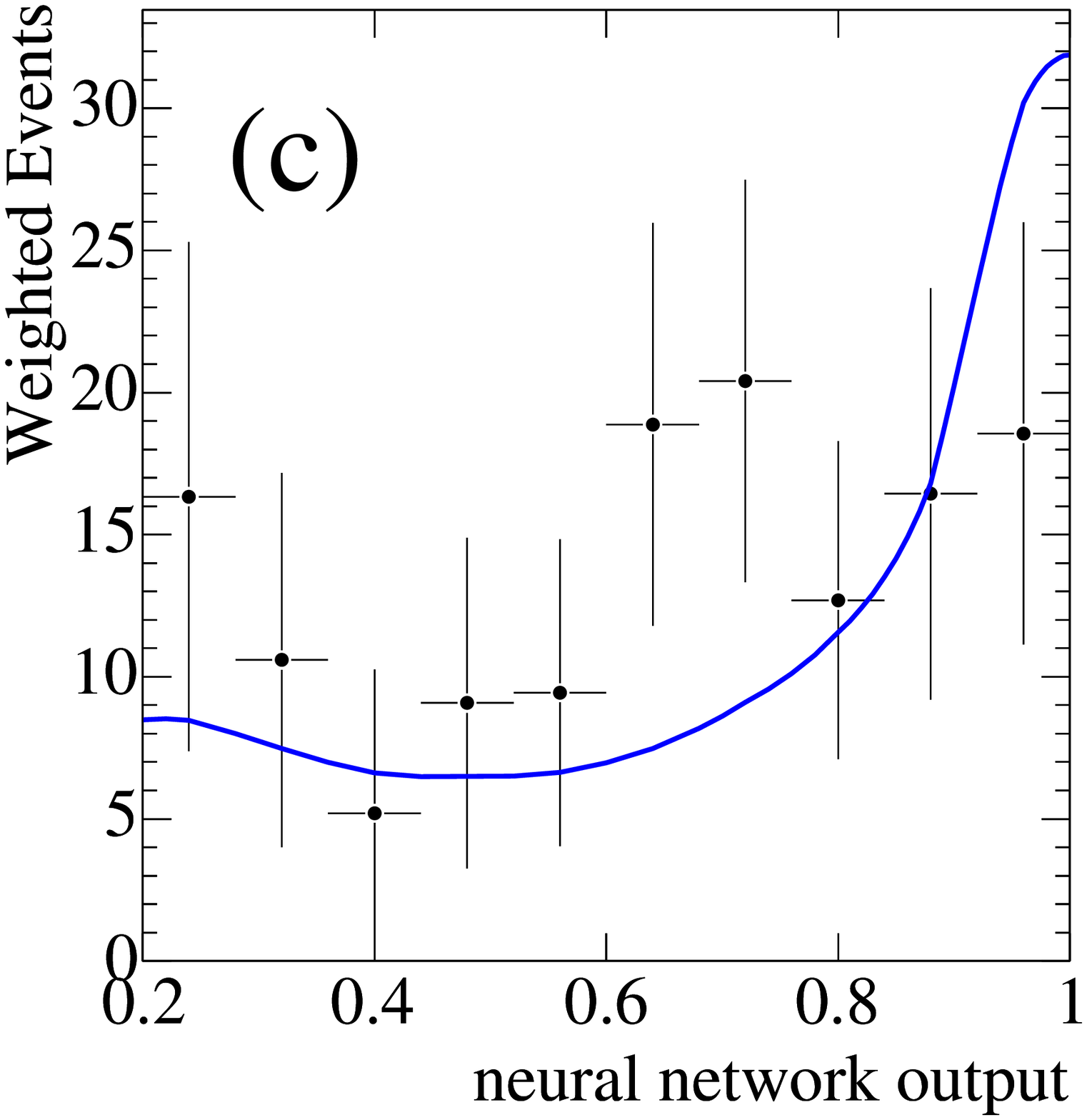}
    \end{minipage}
    &
    \begin{minipage}{1.72in}
      \centering
      \includegraphics[height=1.7in,width=1.7in]
		      {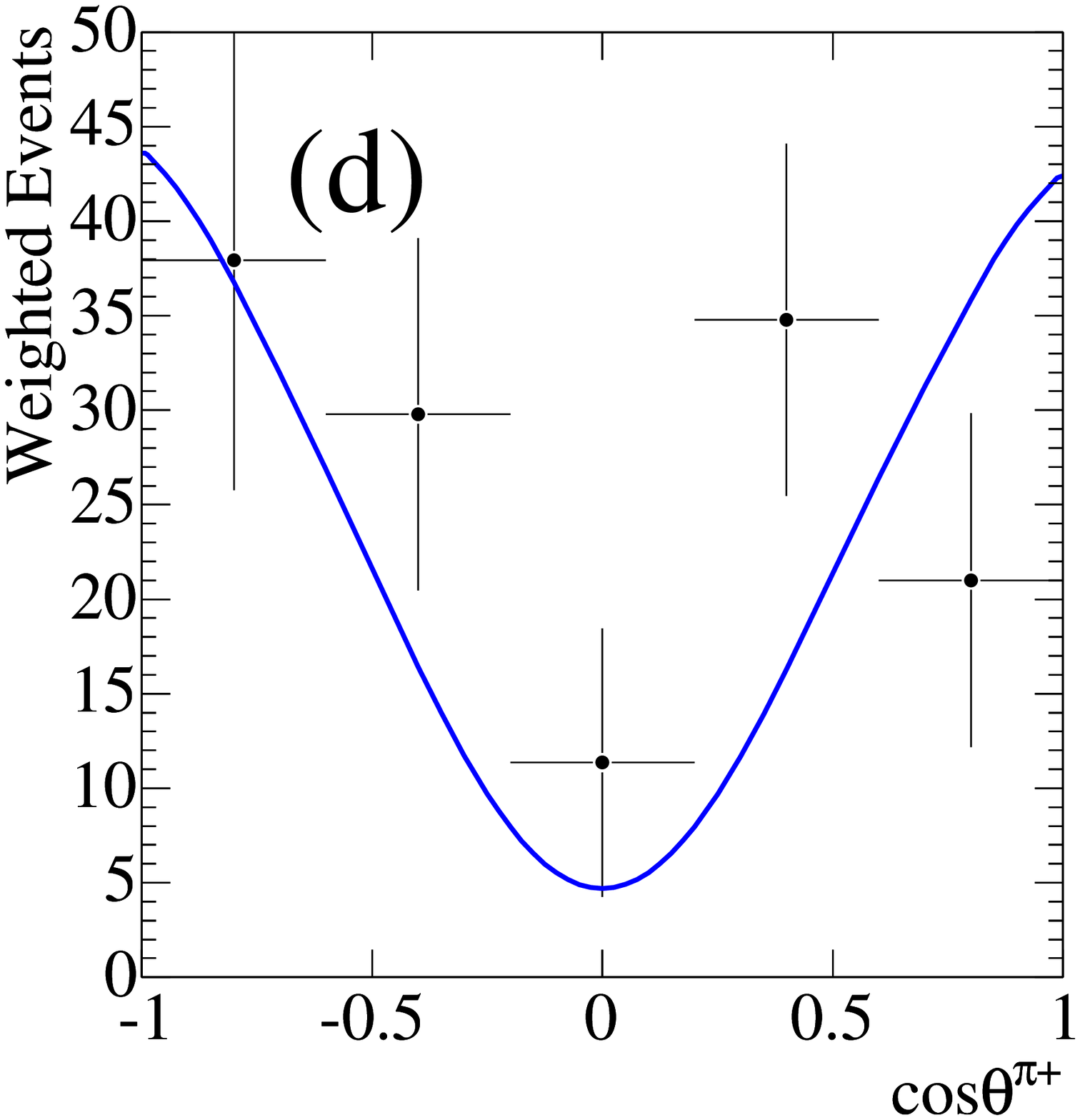}
    \end{minipage}
  \end{tabular}
\centering
\begin{tabular}{cc}
\begin{minipage}{1.72in}
  \centering
   \includegraphics[height=1.7in,width=1.7in]
		  {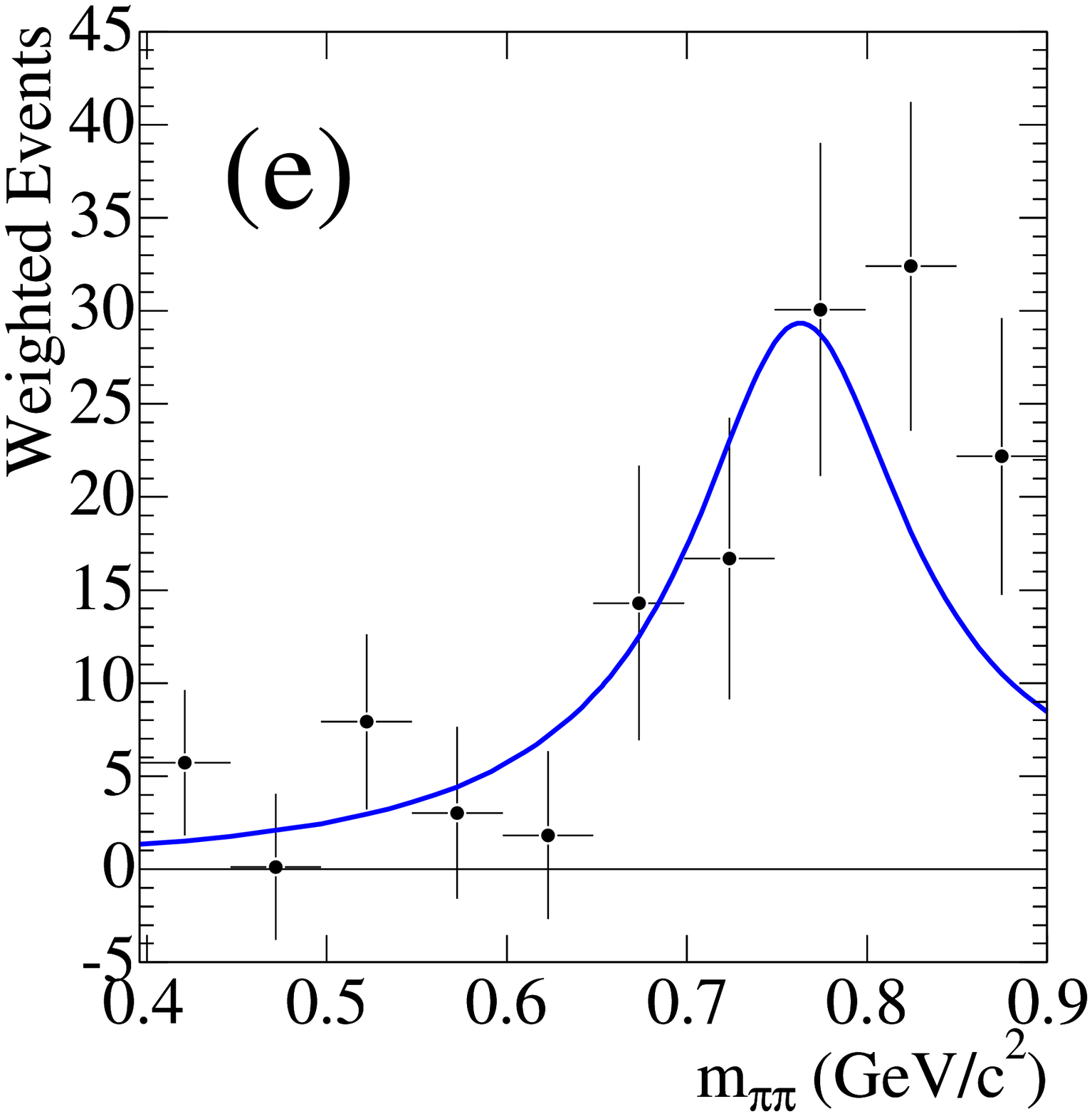}
\end{minipage}
&
\begin{minipage}{1.72in}		  
\centering	  
   \includegraphics[height=1.7in,width=1.7in]
                  {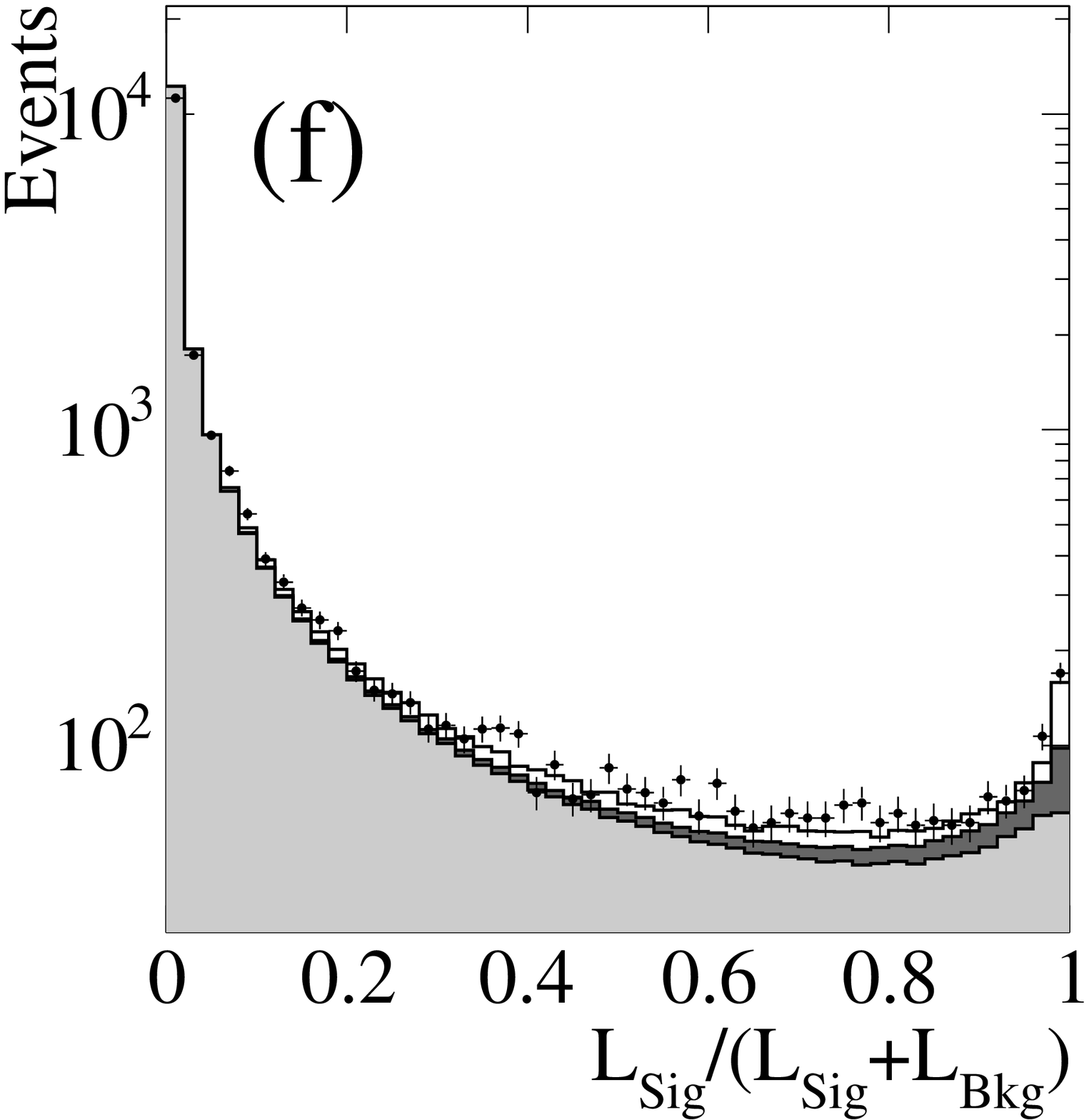}
\end{minipage}	  
  \end{tabular}
    \caption{$_s\cal{P}\it{lots}$ of Maximum Likelihood fit discriminating variables: (a)  $m_{\rm ES}$, (b) $\Delta E$, (c) Neural Network output, (d) $\cos\theta_{\pi^+}$, (e) invariant mass of the $\pi^+\pi^-$ combination. Lines are projections of signal PDFs for each variable. (f) is a plot of the likelihood of an event being signal calculated for all events in our dataset and compared to the predictions of our PDF (predicted continuum in light grey, $B$ Background in dark grey and signal in white).} 
\label{fig:sPlots}
\end{figure}

The $\mes$, $\de$, \NN, $\cos\theta_{\pi^+}$ and $m(\pi^+\pi^-)$  PDFs 
for signal and $B$ background are taken from MC simulation. In general they are non-parametric, with the exception of $\mes$ and $\de$ for signal signal PDFs appear as solid curves in Figure \ref{fig:sPlots}
The $CP$ parameters for $\eta'\KS$ and $f_0 \KS$ backgrounds are fixed to $C = 0$ and 
$S=\sin 2\beta$ (for $\eta'\KS$) and $S=-\sin 2\beta$ (for $f_0\KS$), in accordance with SM expectations. 
For the remaining $B$ backgrounds the parameters $C$ and $S$ are fixed to
$0$.
The PDF parameters describing the continuum background are either allowed to vary freely in the fit or else determined separately from off-resonance data.

There are 16 free parameters in the fit: the yield of signal events, $S_{\rho\KS}$ and $C_{\rho\KS}$ and 13 that parameterize the
continuum background. The continuum parameters are: the yields (5), and those 
asociated with the second order polynomial describing the
$\Delta{E}$ distribution (2), the ARGUS \cite{ARGUS} function
describing the $m_{\rm ES}$ distribution (1) and the double Gaussian used
to model the $\Delta{t}$ distribution (5).

\par
The fit yields $111\pm19$ signal events. We calculate the branching fraction from the measured signal yield, efficiency (including the $\rho^0\to\pip\pim$, $K^0\to\KS$ and $\KS\to\pip\pim$ branching fractions), and the number of $\BB$ events. The result is ${\cal B}(B^0\to\rho^{0}K^0)=(4.9\pm0.8\pm0.9)\times10^{-6}$, where the first error is statistical and the second systematic. The likelihood ratio between the fit result of $111$ signal events and the null hypothesis of zero signal shows that this is excluded at the $8.7\sigma$ level. When additive systematic effects are included we exclude the null hypothesis at the $5.0\sigma$ level.
The fit for $CP$ parameters gives $S_{\rho^{0}\KS}=0.20\pm{0.52}\pm{0.24}$ and $C_{\rho^{0}\KS}=0.64\pm{0.41}\pm{0.20}$.

\par

Figure \ref{fig:sPlots} shows $_s\cal{P}\it{lots}$ ~\cite{sPlot} of the discriminating variables in the fit. Knowledge of the level of background and our ability to distinguish it from signal can be gained from the errors in these plots. In addition, Fig.~\ref{fig:sPlots}(f) shows
the ratio ${\cal L}_S/({\cal L}_S + {\cal L}_B)$ for all events, where
${\cal L}_S$ and ${\cal L}_B$ are the likelihoods for each event to be
signal or background, respectively.  

\par
 
\begin{figure}
  \centering
      \begin{minipage}{3.0in}
      \centering
        \includegraphics[height=2.9in,width=2.9in]
		      {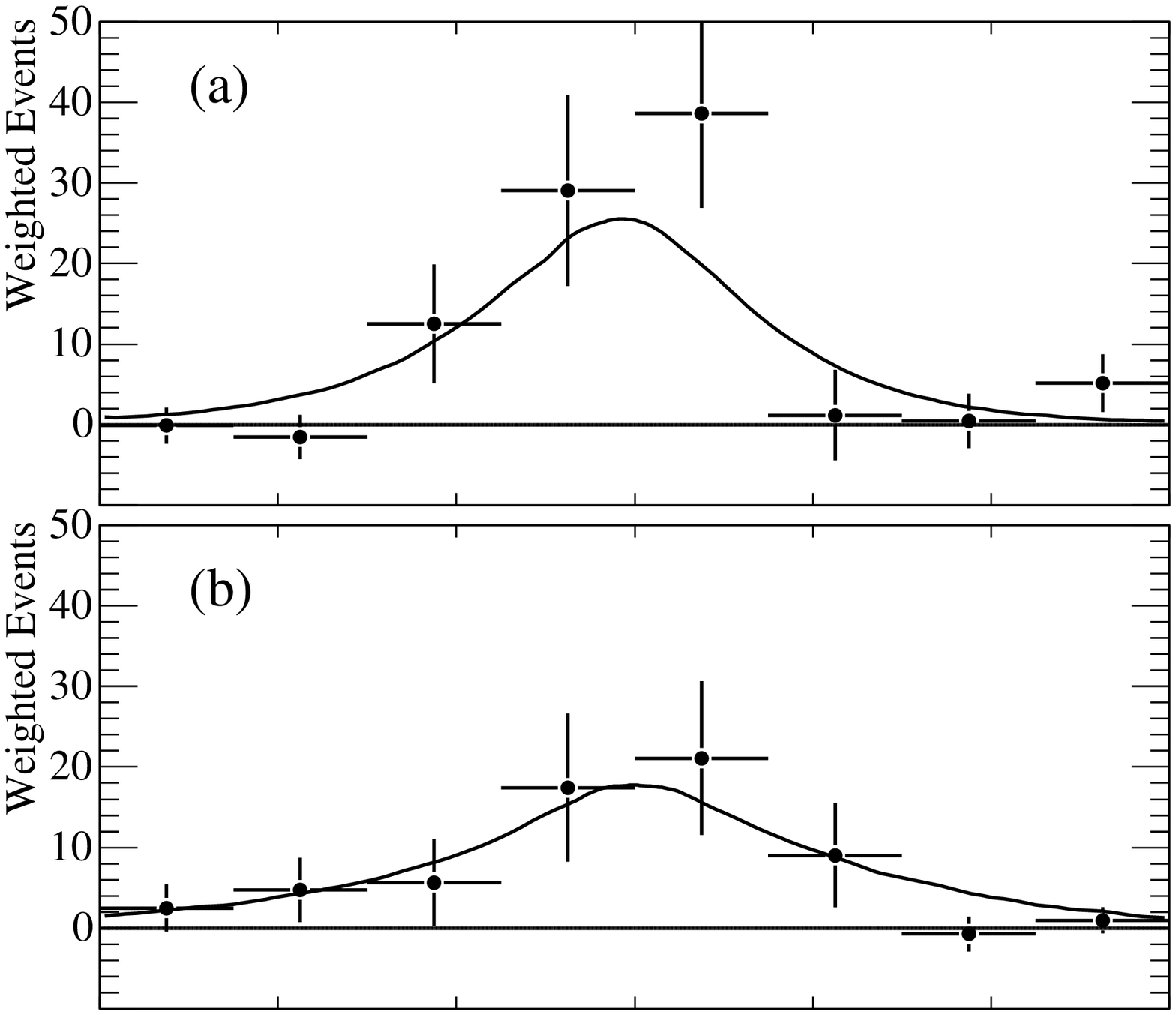}
           \end{minipage}
		      \begin{minipage}{3.0in}
		        \centering
\includegraphics[height=1.64in,width=2.9in]
                      {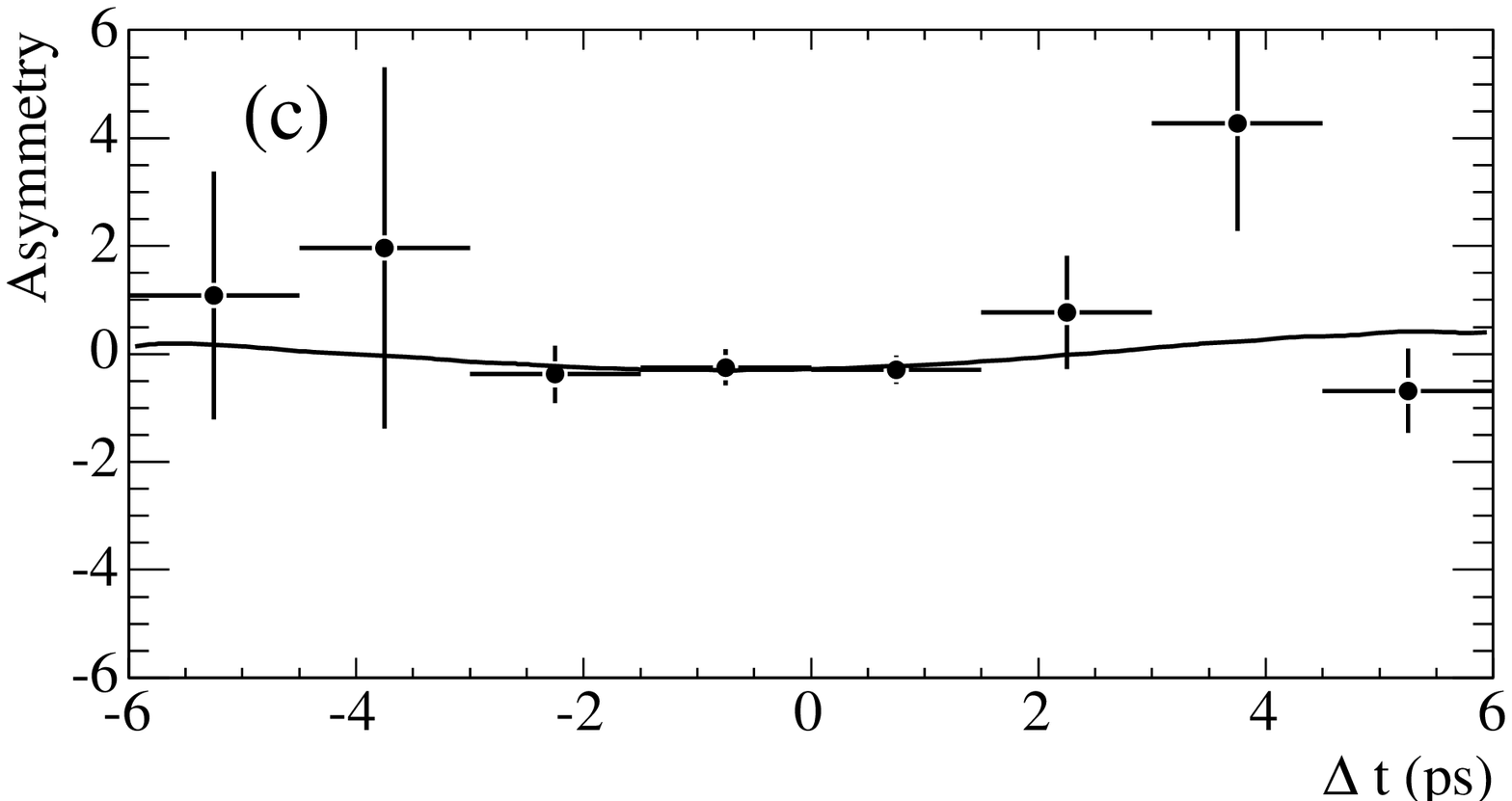}
                      \end{minipage}
		       \caption{$_s\cal{P}\it{lots}$ of $\Delta{t}$, overlaid with projected signal PDFs, split into (a) $\Bz_{\rm tag}$ tags, (b) $\Bzb_{\rm tag}$ tags and (c) the asymmetry $(N_{\Bz_{\rm tag}}-N_{\Bzb_{\rm tag}})$$/(N_{\Bz_{\rm tag}}+N_{\Bzb_{\rm tag}})$ as a function of $\Delta{t}$ .}
\label{fig:dtPlots}
\end{figure}

Figure \ref{fig:dtPlots} shows $_s\cal{P}\it{lots}$ of $\Delta{t}$. Untagged events are removed, and events are split into $\Bz_{\rm tag}$ tags and $\Bzb_{\rm tag}$ tags. An $_s\cal{P}\it{lot}$ of asymmetry 
$(N_{\Bz_{\rm tag}}-N_{\Bzb_{\rm tag}})/(N_{\Bz_{\rm tag}}+N_{\Bzb_{\rm tag}})$ as a function of
 $\Delta{t}$ is also shown.

\par
Systematic errors are listed in Table \ref{tab:systematics}.
We estimate biases due to the fit procedure from fits to a large number of simulated experiments. We vary parameters fixed in the nominal fit by their uncertainty and include the change in result as the corresponding systematic error. 
The systematic uncertainties arise from sources including the parameterization of the signal $\Delta t$ resolution
function, the mistag fractions, and discrepancies between data and the simulation including the effect of alternative models for resonances. 

We estimate the systematic uncertainly due to neglecting the intereference between $\Bz\to\KS\pi^+\pi^-$ from both parameterized and full simulations that take interference into account. We include contibutions from $\rho^0(770) \KS$, $f_0(980)\KS$,
$K_0^*(1430)^+\pi^-$, $K_0^*(892)^+\pi^-$ and $f_2(1270)\KS$, as well as two $\KS\pi\pi$ non-resonant contributions. 
We simulate many samples with different relative phases between modes. We also vary the amplitude of each mode within limits based on the best available information \cite{bellekpipi,babarkpipi}. Each simulation is then subjected to the standard selection and fitting procedure. The systematic uncertainty is taken from the width of a Gaussian fitted to the distribution of the results.

\begin{table}
\centering
\begin{tabular}{lccc}
\hline \hline
\hline 
Mis-reco'd events and fit bias   & 0.12 &0.09 & 10\\
PDF uncertainties               & 0.13 &0.18 & 2\\
Tagging parameters              & 0.02 &0.01 & -\\
Neglect of intereference        & 0.14&0.09 & 7\\
$\rho^0$ mass shape             & 0.07& 0.05 & 3\\
$B$ Background BF                 & 0.02&0.10 & 13\\
$CP$ of background              & 0.04& 0.00 & - \\
Tracking efficiency \& $B$ counting             &  -   & -    & 6\\
\hline
Total &0.24  & 0.20 & 19 \\
\hline \hline
\end{tabular}
\caption{Summary of contributions to the systematic error.}
\label{tab:systematics}
\end{table}

In summary, we have established the existance of the decay	
$\Bz\to\rho^0K^0$ and measured its branching fraction with the	
significance of 5 standard deviations. Our measurement agrees within errors 
with ${\cal B}(\Bz\ra\omega\Kz)$ as measured in~\cite{bbOmK0}, as 
expected if a single 
penguin amplitude dominates these decays. We have extracted the $CP$ violating
 parameters $S$ and $C$ for $\Bz\to\rho^0\KS$ 
which are consistent with those measured in charmonium channels \cite{Aubert:2004zt}.

We are grateful for the excellent luminosity and machine conditions
provided by our \pep2\ colleagues, 
and for the substantial dedicated effort from
the computing organizations that support \babar.
The collaborating institutions wish to thank 
SLAC for its support and kind hospitality. 
This work is supported by
DOE
and NSF (USA),
NSERC (Canada),
IHEP (China),
CEA and
CNRS-IN2P3
(France),
BMBF and DFG
(Germany),
INFN (Italy),
FOM (The Netherlands),
NFR (Norway),
MIST (Russia),
MEC (Spain), and
PPARC (United Kingdom). 
Individuals have received support from the
Marie Curie EIF (European Union) and
the A.~P.~Sloan Foundation.

\end{document}